\documentclass[amssymb,prb,twocolumn,showpacs]{revtex4}
\input{psfig.sty}
\usepackage{epsfig}
\usepackage{dcolumn}
\usepackage{amsmath}
\begin{document}
%\documentclass[12 pt,a4paper]{article} %selecciona el tipo de documento
%\usepackage[english]{babel} %selecciona el idioma
%\frenchspacing %trata los espacios despues de los puntos igual que los otros
%\usepackage{epsfig}
%\usepackage{amsmath}
%\usepackage[a4paper,dvips]{geometry}
%\geometry{textwidth=16 cm, textheight=22 cm}
%\begin{document}
\title{\bf\ From zero resistance states to absolute negative
conductivity in microwave irradiated 2D electron systems}
\author{J. I\~narrea$^{1,2}$
 and G. Platero$^1$}
\affiliation{$^1$Instituto de Ciencia de Materiales,
CSIC, Cantoblanco, Madrid, 28049, Spain \\
$^2$Escuela Polit\'ecnica Superior, Universidad Carlos III, Leganes,
Madrid, 28911, Spain}
\date{\today}
%%%%%%%%%%%%%%%%%%%%%%%%%%%%%%%%%%%%%%%%%%%%%%%%%%%%%%%%%%%%%%%%%%%%%%%%%%%%%%
\begin{abstract}
Recent experimental results regarding a 2D electron gas subjected to
microwave radiation reveal that magnetoresistivity, apart from
presenting oscillations and zero resistance states,  can evolve to
negative values at minima. In other words, the current can evolve
from flowing with no dissipation, to flow in the opposite direction
of the dc bias applied. Here we present a theoretical model in which
the existence of radiation-induced absolute negative conductivity is
analyzed. Our model explains the transition from zero resistance
states to absolute negative conductivity in terms of multiphoton
assisted electron scattering due to charged impurities and shows how
this transition can be driven by tuning microwave frequency and
intensity. This opens the possibility of controlling the
magnetoconductivity in microwave driven nanodevices and
understanding the novel optical and transport properties of such
devices.
\end{abstract}
%%%%%%%%%%%%%%%%%%%%%%%%%%%%%%%%%%%%%%%%%%%%%%%%%%%%%%%%%%%%%%%%%%%%%%%%%%%%%%
\maketitle

The effect of an AC field on the electronic transport properties of
nanostructures has been an active research topic in the last years.
One reason for this is that the AC field can profoundly modify the
electronic structure and the dynamical properties of electrons in
the nanostructure. The application of an external AC potential also
allows the electronic properties to be tuned in a controllable
way\cite{grossmann,lopez}. Ten years ago, transport experiments on
AC-driven weakly coupled semiconductors superlattices reveled a
fascinating, non-intuitive, behavior: for certain parameters of the
AC potential and stationary electric field, the electronic current
flowed $uphill$ presenting absolute negative conductivity
(ANC)\cite{keay}. In the rapid developing field of nanoelectronics,
it can be expected that nanodevices, will soon routinely incorporate
two-dimensional electron systems. A deep knowledge of how this
structures respond to external electromagnetic fields is thus of
high importance.

Recently two experimental groups\cite{mani,zudov} have announced
the existence of oscillations and zero resistance states (ZRS) in
the longitudinal magnetoresistivity ($\rho_{xx}$) of two
dimensional electron systems (2DES) subjected to microwave (MW)
radiation and a perpendicular magnetic field ($B$). One of the
most controversial topics in this field, has been the existence of
ANC and even the very presence of ZRS has been also questioned.
Most experimental
works\cite{mani,zudov,mani2,mani3,potemski1,yang,du,smet}, report
clearly such vanishing dissipation states but not ANC. Only
Willett et al\cite{willett}, and very recently Zudov et
al\cite{zudov2}, have reported minima in which $\rho_{xx}$ is
distinctly negative. On the other hand many theoretical
contributions have been presented to explain $\rho_{xx}$
oscillations with $B$ and the possibility of ZRS and
ANC\cite{girvin,dietel,lei,lei2,ryzhi,rivera,shi,ahn,andreev,ina,ng,ina2}.

Here we present a microscopical model which explains how the system
evolves from ZRS to ANC. The proposed theory is based on how the
Larmor orbits dynamics are driven by the MW field and the external
DC bias ($E_{DC}$) in the transport direction ($x$-direction). It
shows how this transition can be driven by tuning the MW parameters.
We start from an exact expression for the electronic wave function
dressed by photons. This many body wave function is expressed in
terms of photosatellites and energy sidebands\cite{ina3,ina4}. At
high MW-power and low frequency, multi-photon processes become
relevant in assisting electron-charged impurity scattering,
responsible for $\rho_{xx}$. This exact solution for the electronic
wave function of a 2DES in a perpendicular $B$-field, a DC electric
field and MW radiation, is given by \cite{ina,kerner,park}:
\begin{eqnarray}
\Psi(x,t)\propto\phi_{n}(x-X-x_{cl}(t),t)
%&&\times  exp
%\left[i\frac{m^{*}}{\hbar}\frac{dx_{cl}(t)}{dt}[x-x_{cl}(t)]+
%\frac{i}{\hbar}\int_{0}^{t} {\it L} dt'\right]\nonumber  \\
\sum_{p=-\infty}^{\infty} J_{p}\left[\propto \frac{eE_{0}X }{\hbar
w} \right]
%\left(\frac{1}{w}+\frac{w}{\sqrt{(w_{c}^{2}-w^{2})^{2}+\gamma^{4}}}\right)
e^{ipwt}
\end{eqnarray}
where $e$ is the electron charge, $\phi_{n}$ is the solution for the
Schr\"{o}dinger equation of the unforced quantum harmonic
oscillator, $w$ the MW frequency, $E_{0}$ the intensity for the MW
field, $X$ is the center of the orbit for the electron motion,
$x_{cl}=A\cos wt$ where $A\propto\frac{e E_{o}}{m^{*}w^{2}}$ and
$J_{p}$ are Bessel functions. The wave function includes the energy
sidebands $\epsilon_{n}$,$\epsilon_{n} \pm \hbar w$,$\epsilon_{n}\pm
2 \hbar w$,....$\epsilon_{n} \pm p \hbar w $ and shows that, due to
the MW radiation, the centers of electronic orbits are not fixed,
but instead oscillate back and forth harmonically with $w$ and
amplitude $A$. Electrons suffer scattering due to charged impurities
that are randomly distributed in the sample. To proceed we calculate
the electron-charged impurity multi-photon assisted transition rate
$W_{n,m}$, from an initial state $\Psi_{n}(x,t)$, to a final state
$\Psi_{m}(x,t)$\cite{ridley}:
%\begin{equation}
%W_{n,m}= \frac{e^{5}n_{i}B S} {16\pi^{2}\hbar^{2}\epsilon^{2}}\int
%dq \frac{q}{(q^{2}+q_{0}^{2})^{2}} \frac{n_{1}!}{n_{2}!}
%e^{-\frac{1}{2}q^{2}R^{2}}&& \nonumber\\
%\times
%\left(\frac{1}{2}q^{2}R^{2}\right)^{n_{1}-n_{2}}\left[L_{n_{2}}^{n_{1}-n_{2}}(
%\frac{1}{2}q^{2}R^{2})\right]^{2}
$W_{n,m}=W_{n,m}(0)[B_{0}+B_{1}+B_{2}]$
%\end{equation}
where, $W_{n,m}(0)$ is the transition rate when $J_{0}\simeq 1 $,
and
\begin{eqnarray}
&&B_{0}=[J_{0}^{2}(A_{m})J_{0}^{2}(A_{n})+\sum_{1}^{s}2J_{s}^{2}(A_{m})
J_{s}^{2}(A_{n})] \nonumber\\
&&\times \left[\frac{\Gamma}{[\hbar
w_{c}(n-m)]^{2}+\Gamma^{2}}\right]\\
%\end{eqnarray}
%\begin{eqnarray}
&&B_{1}=[\sum_{0}^{s}J_{s}^{2}(A_{m})J_{s+1}^{2}(A_{n})+J_{s+1}^{2}(A_{m})
J_{s}^{2}(A_{n})]\nonumber\\
&&\times \left[\frac{\Gamma}{[\hbar w_{c}(n-m)+\hbar
w]^{2}+\Gamma^{2}}\right]
\end{eqnarray}
\begin{eqnarray}
&&B_{2}=[J_{1}^{2}(A_{m})J_{1}^{2}(A_{n})+
\sum_{0}^{s}J_{s}^{2}(A_{m})J_{s+2}^{2}(A_{n})+ \nonumber\\
&&J_{s+2}^{2}(A_{m})J_{s}^{2}(A_{n})]
\times\left[\frac{\Gamma}{[\hbar w_{c}(n-m)+2\hbar
w]^{2}+\Gamma^{2}}\right]
\end{eqnarray}
Here $A_{i}\propto\frac{E_{0}}{w}$\cite{ina} and $\Gamma$ is the
state (Landau level) broadening due to different scattering
mechanisms. Multiphoton processes have been considered up to 2
photons.

The average effective distance advanced by the electron in every
scattering jump is given by: $\Delta X^{MW}=\Delta X^{0}+ A\cos
w\tau$, where $\Delta X^{0}$ is the effective distance advanced when
there is no MW field present and $1/\tau=W_{n,m}$ ($\tau$ being the
impurity scattering time). From here, the longitudinal conductivity
$\sigma_{xx}$ can be calculated through $\sigma_{xx}\propto \int
\rho(E_{n})\frac{\Delta X^{MW}}{\tau}(f_{i}-f_{f})dE_{n}$, where
$f_{i}$ and $f_{f}$ are the distribution functions for the initial
and final Landau states respectively:
\begin{eqnarray}
&&\sigma_{xx}=\sigma_{xx}(0)[B_{0}[f(E_{n})-f(E_{m})]+\nonumber\\
&&B_{1}[f(E_{n})-f(E_{m}+\hbar w)]+ B_{2}[f(E_{n})-f(E_{m}+2\hbar
w)]] \nonumber\\
\end{eqnarray}
and $\sigma_{xx}(0)$ is the conductivity when $J_{0}\simeq 1 $. To
obtain $\rho_{xx}$ we use the relation
$\rho_{xx}=\frac{\sigma_{xx}}{\sigma_{xx}^{2}+\sigma_{xy}^{2}}
\simeq\frac{\sigma_{xx}}{\sigma_{xy}^{2}}$, where
$\sigma_{xy}\simeq\frac{n_{i}e}{B}$ and $\sigma_{xx}\ll\sigma_{xy}$.
\begin{figure}
\centering \epsfxsize=3.5in \epsfysize=3.0in \epsffile{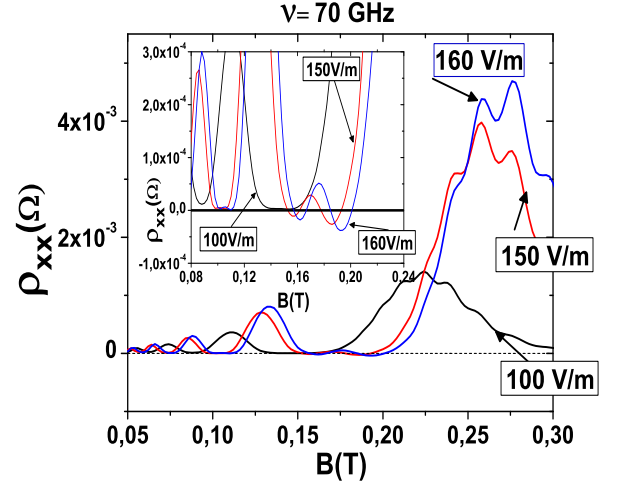}
\caption{Calculated magnetoresistivity $\rho_{xx}$ as a function of
$B$, for different MW-intensities at $\nu=70 GHz$. In the inset we
show amplifications of principal minima.  Zero resistance states,
absolute negative conductivity and the corresponding evolution for
increasing MW-power can be observed Temperature: 500 mK.}
\end{figure}

In Fig. 1 the calculated $\rho_{xx}$ as a function of $B$ for
different MW power at fixed frequency $\nu=w/2\pi=70 GHz$ is shown.
Our results follow the qualitative experimental
behavior\cite{willett}, and show that ANC occurs at the principal
minimum, as the incident radiation power is increased.
 The inset shows the evolution from ZRS
to negative conductivity. The physical explanation is as follows. In
Fig. 2 we represent schematic diagrams to describe $\rho_{xx}$
evolution at minima. In Fig. 2a orbits are moving forwards and on
average the electron advances a shorter distance than in the no MW
case, $\Delta X^{MW}<\Delta X^{0}$. This corresponds to a decrease
in the conductivity but $\rho_{xx}>0$ still. If we raise the MW
power we will eventually reach the situation depicted in Fig. 2b,
where orbits are moving forwards but their amplitude $A$ is larger
than the electronic jump. In this case the jump is blocked by the
Pauli exclusion principle because the final state is occupied. {\it
This is the physical origin of the ZRS }.
\begin{figure}
\centering\epsfxsize=3.5in \epsfysize=3.5in \epsffile{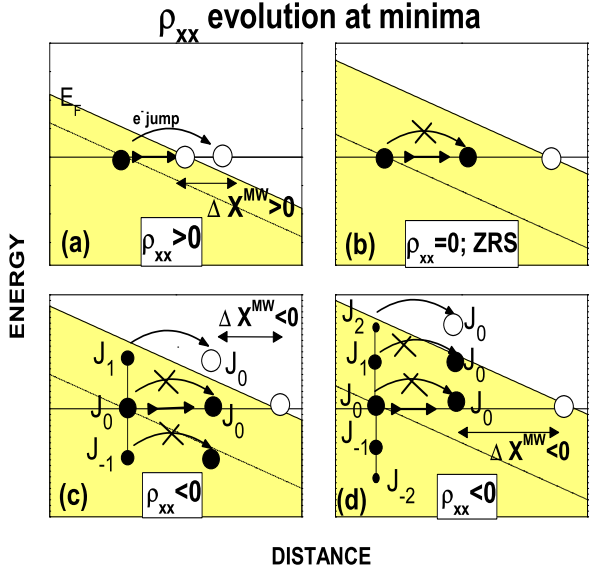}
\caption{Schematic diagrams of electronic transport describing
$\rho_{xx}$ evolution at minima, for fixed MW frequency and
different MW intensities. When the MW-field is on, electronic orbits
oscillate with $w$ and at minima, displacement is forwards. In (a)
the average distance advanced by the electron $\Delta X^{MW}$ is
shorter than the no MW case $\Delta X^{0}$($\Delta X^{MW}<\Delta
X^{0}$). This corresponds to a reduction in the conductivity but
still $\rho_{xx}>0$. In (b), at higher intensity we can reach a
situation where the amplitude $A$, of the electronic orbit
oscillations, is larger than electronic jump, and the electronic
movement between orbits cannot take place because the final state is
occupied. This situation corresponds to ZRS and $\rho_{xx}\sim 0$.
In (c) and (d) due to further increases in MW-power the wave
function sidebands play and important role. In (c) one photon,
$J_{1}\rightarrow J_{0}$, transitions can take place giving rise to
negative conductivity ($\Delta X^{MW}<0$ and
$(f_{i}-f_{f})>0\Rightarrow \rho_{xx}<0$). In (d) the negative
contribution transition corresponds to two photons,
$J_{2}\rightarrow J_{0}$, transitions also giving $\rho_{xx}<0$. }
\end{figure}
\begin{figure}
\centering\epsfxsize=3.5in \epsfysize=3.0in \epsffile{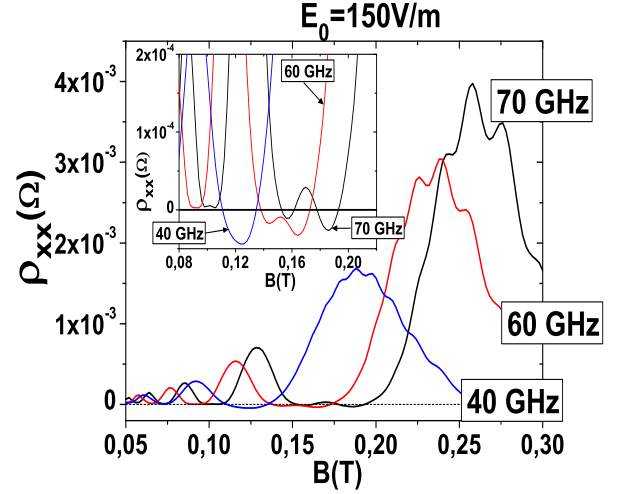}
\caption{Calculated magnetoresistivity $\rho_{xx}$ as a function of
$B$, for different frequencies at MW-intensity $E_{0}=150V/m$. In
the inset we display a rising central peak in the middle of negative
minimum that becomes positive as the MW-frequency is increased.
Temperature: 500 mK.}
\end{figure}
At fixed photon frequency, small values for MW-power correspond to
transitions where no photon absorption or emission are involved,
i.e., the arguments of Bessel functions are so small that only
$J_{0}$ terms need to be taken into account. This implies that only
direct, $J_{0}\rightarrow J_{0}$, transitions are relevant. This
corresponds to Fig. 2a and 2b. If we further increase the MW-power,
while keeping the frequency constant, the argument of the Bessel
functions will become larger, and higher order sidebands must be
considered: multi-photon transitions become relevant. Transitions
such  $J_{1}\rightarrow J_{0}$ and $J_{2}\rightarrow J_{0}$, which
correspond to one photon and two photons processes respectively, can
then play and effective role in the current. At minima and for $A$
larger than the electron scattering jump (large MW-power), we find a
situation where $\Delta X^{MW}<0$. However for multi-photon
transitions $J_{1}\rightarrow J_{0}$ or $J_{2}\rightarrow J_{0}$
etc, the difference of distribution functions $(f_{i}-f_{f})>0$.
These processes ($\Delta X^{MW}<0$ and $(f_{i}-f_{f})>0 \Rightarrow
\rho_{xx}<0$) produce negative contributions to the current and {\it
are the physical origin of ANC}.

One surprising effect in the experimental results is the positive
peak in the middle of $\rho_{xx}$ negative minimum. This is also
observed in calculated results. (see inset of Fig. 1). The
explanation for this can be readily obtained. In a regime with
$\Delta X^{MW}<0$ and finite temperature, direct ($J_{0}\rightarrow
J_{0}$) transitions, can correspond to negative values for the
difference of electronic distribution functions, $(f_{i}-f_{f})$.
This is due to the fact that in such a regime, the final state is
always deeper in energy than the initial state, with respect to the
Fermi energy. Considering the smearing of the distribution function
at finite temperature, the final result is that $f_{i}<f_{f}
\Rightarrow (f_{i}-f_{f})<0$.

When  $\Delta X^{MW}<0$ and $(f_{i}-f_{f})<0$, an effective positive
net current will be produced giving rise to a positive $\rho_{xx}$.
In the inset of Fig. 1 it can also be observed that an increase in
MW-power produces an increase in the positive and negative part of
the minimum in a similar way. The explanation has to do with the
corresponding increase in the amplitude $A$ and, as a consequence,
in $\Delta X^{MW}$. This has a similar impact on both the
one-photon, $J_{1}\rightarrow J_{0}$, transitions (negative
contributions) and on the direct $J_{0}\rightarrow J_{0}$ ones,
(positive contributions). Positive net values for $\rho_{xx}$ in the
middle of the main minima, have been experimentally obtained by
other groups\cite{mani3,willett}, notably in ref [7], where this
effect was termed "breakdown of ZRS".
\begin{figure}
\centering \epsfxsize=3.5in \epsfysize=3.5in \epsffile{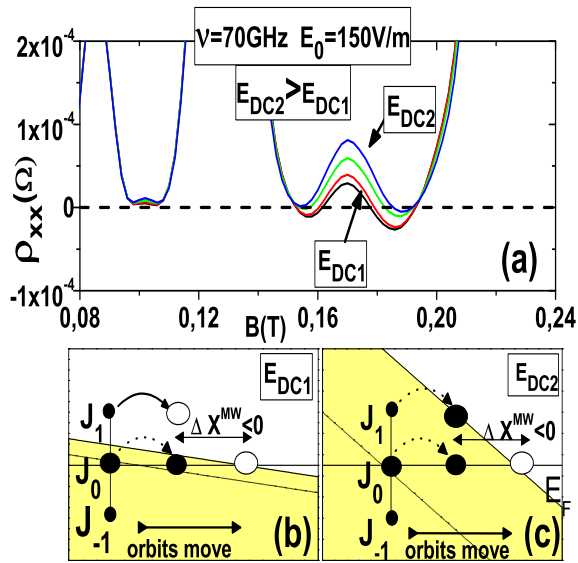}
\caption{(a) Calculated $\rho_{xx}$ versus $B$ at two minima for
fixed MW-frequency and intensity and different external DC bias
($E_{DC}$). For increasing $E_{DC}$, $\rho_{xx}$ minimum becomes
positive. (b) Schematic diagrams explanatory for the lower bias case
($E_{DC1}$). In this situation the Fermi energy slope is smaller
making the $J_{1}\rightarrow J_{0}$ transitions predominant which
provides negative contributions to the current. (c) Schematic
diagram for higher bias ($E_{DC2}$). In this case the former
$J_{1}\rightarrow J_{0}$ negative contributions become positive
because the final Landau state is now below the Fermi energy and
therefore is not empty. As a result, most contributions are positive
shifting the minimum of $\rho_{xx}$ upwards and causing the central
peak to increase. Temperature 500 mK.}
\end{figure}

Fig. 3 shows $\rho_{xx}$ versus $B$ for different MW-frequencies at
fixed MW-power. ANC is achieved for all the cases presented and the
growth of the central peak as frequency is increased is clearly
visible (see inset). The explanation is as follows. At constant
$E_{0}$, lower $w$ corresponds to high values for the Bessel
function arguments, i.e., $J_{0}$'s are decreasing and $J_{n\neq
0}$'s are increasing. In this regime, direct transitions (positive
contributions) become less important than multi-photon,
$J_{n}\rightarrow J_{0}$ transitions (negative contributions). The
first ones are represented by the $B_{0}$ term in the scattering
rate $W_{n,m}$ and $\sigma_{xx}$, and the second ones by $B_{1}$ and
$B_{2}$, (see Eqs. 2-5). Eventually the positive contributions are
totally compensated by the negative ones and the central peak at
minimum vanishes. As we increase the frequency, the reverse, occurs
giving rise to a distinct positive peak in the middle of the
negative minimum.

According to the model and calculated results presented above, high
values for the ratio $\frac{E_{0}}{w}$  is a first and important
condition to obtain photo-satellites and the consequent ANC. However
this is not sufficient, and an additional condition has to be
fulfilled concerning the external DC bias applied in the $x$
direction, as we explain below.

In Fig. 4a we represent calculated $\rho_{xx}$ versus $B$ at fixed
$w$ and $E_{0}$ and at increasing external DC bias $E_{DC}$. We can
observe at the two main minima how the $\rho_{xx}$ negative values
are shifted to positive as $E_{DC}$ is raised. At the same time the
central peak gets larger. The explanation can be seen from the
schematic diagrams in Fig. 4b and 4c. In both cases we are in a
regime where $\Delta X^{MW}<0$. At lower bias ($E_{DC1}$, Fig 4.b),
one-photon transitions giving negative current ($J_{1}\rightarrow
J_{0}$) dominate and, except in the central peak, they can
compensate the positive direct ($J_{0}\rightarrow J_{0}$)
contributions. We observe that the final state for $J_{1}\rightarrow
J_{0}$ is well above the Fermi energy and the corresponding
difference $(f_{i}-f_{f})>0$, is important making the negative
contributions relevant. At higher bias $E_{DC2}$, we reach the
situation depicted in Fig. 4c. The corresponding slope for the Fermi
energy is larger than in the $E_{DC1}$ case. Now the final state is
below $E_{F}$, i.e., it is not empty making the corresponding
difference $(f_{i}-f_{f})<0$ and the contribution to the current
evolves from negative to positive. The final result is that the
$\rho_{xx}$ negative values get smaller for increasing $E_{DC}$
becoming eventually zero or positive. Another important results is
that the central $\rho_{xx}$ peak becomes much larger which can be
explained with similar arguments.

Therefore we can state that high values for $\frac{E_{0}}{w}$ and a
regime of lower values for  $E_{DC}$, constitute the two main
conditions which have to be fulfilled in order to obtain ANC. Thus
according to our theoretical model and taking into account of the
experimental difficulties to measure negative resistance, we believe
that by tuning the appropriate MW  parameters and external DC bias,
it would be possible to experimentally observe the evolution from
ZRS to ANC.

In summary, we have presented here a theoretical model to explain
the physics behind the existence of radiation-induced ANC in 2D
electron systems. We are able to explain the transition from ZRS to
ANC in terms of multi-photon assisted processes. It has also been
shown how this transition can be tuned by the parameters of the
microwave field. In addition, the external DC bias has been found to
be of decisive importance in achieving ANC. We think that this work
sheds some light on the controversy whether ANC actually occurs in
these structures and which are the main parameters that govern the
evolution from ZRS to ANC. The basic understanding of this process
will be of importance to the future implementation of workable
nanodevices.

We acknowledge Charles Creffield for the critical reading of the
manuscript. This work was supported by the MCYT (Spain) grant
MAT2002-02465, the ``Ramon y Cajal'' program (J.I.). and the EU
Human Potential Programme HPRN-CT-2000-00144.


\begin{thebibliography}{20}
\bibitem{grossmann}
F.Grossmann, T.Dittrich, P.Jung and P. Hanggi, Phys. Rev. Lett. {\bf
67},516, (1991).
\bibitem{lopez}
R Lopez, R.Aguado, G.Platero and C.Tejedor, Phys. Rev. B, {\bf
64},075319, (2001).
\bibitem{keay}
B.J. Keay et al., Phys. Rev. Lett. {\bf 75},4098, (1995).
\bibitem{mani}
R.G. Mani, J.H. Smet, K. von Klitzing, V. Narayanamurti, W.B.
Johnson, V. Umansky, Nature {\bf 420} 646 (2002).
\bibitem{zudov}
M.A. Zudov, R.R. Du, N. Pfeiffer, K.W. West, Phys. Rev. Lett. {\bf
90} 046807 (2003).
\bibitem{mani2}
R.G. Mani, V. Narayanamurti, K. von Klitzing, J.H. Smet,  W.B.
Johnson, V. Umansky, Phys. Rev. B {\bf 69} 161306(R) (2004).
\bibitem{mani3}
R.G. Mani, Appl. Phys. Lett. {\bf 85}, 4962, (2004); R.G. Mani,
Physica E, {\bf 22}, 1, (2004); R.G. Mani, Physica E, {\bf 25},
189, (2004).
\bibitem{potemski1}
S.A. Studenikin, M. Potemski, A. Sachrajda, M. Hilke, L.N.
Pfeiffer, K.W. West, Phys. Rev. B, {\bf 71}, 245313, (2005).; S.A.
Studenikin, M. Potemski, P.T. Coleridge, A. Sachrajda, Z.R.
Wasilewski, Solid State Comm {\bf 129}, 341 (2004).
\bibitem{yang}
C.L. Yang, M.A. Zudov, T.A. Knuuttila, R.R. Du, L.N. Pfeiffer and
K.W. West, Phys. Rev. Lett. 91, 096803, (2003);  M.A. Zudov, Phys.
Rev. B, {\bf 69}, 041304, (2003).
\bibitem{du}
R.R. Du, M.A. Zudov, C.L. Yang, L.N. Pfeiffer and K.W. West,
Physica E, {\bf 22}, 7 (2004);   R.R. Du, M.A. Zudov, C.L. Yang,
Z.Q. Yuan, L.N. Pfeiffer and K.W. West, J. Mod. Phys. B, {\bf 18},
3465, (2004).
\bibitem{smet}
J.H.Smet, B. Gorshunov, C.Jiang, L.Pfeiffer, K.West, V. Umansky,
M. Dressel, R. Dressel, R. Meisels, F.Kuchar, and K.von Klitzing,
Phys. Rev. Lett. {\bf 95}, 116804 (2005).
\bibitem{willett}
R.L. Willett, L.N. Pfeiffer and K.W. West, Phys. Rev. Lett. {\bf 93}
026804 (2004).
\bibitem{zudov2}
M.A. Zudov, R.R. Du, L.N. Pfeiffer and K.W. West, Phys. Rev. B,
73, 041303 (2006).
\bibitem{girvin}
A.C. Durst, S. Sachdev, N. Read, S.M. Girvin, Phys. Rev. Lett.{\bf
91} 086803 (2003)
\bibitem{dietel}
C.Joas, J.Dietel and F. von Oppen, Phys. Rev. B {\bf 72}, 165323,
(2005); J.Dietel, L.J. Glazman, F.W.J. Hekking and F. von Oppen
Phys. Rev. B {\bf 71} 045329 (2005).
\bibitem{lei}
X.L. Lei, S.Y. Liu, Phys. Rev. Lett.{\bf 91}, 226805 (2003);
\bibitem{lei2}
X.L. Lei, S.Y. Liu, Phys. Rev. B {\bf 72}, 075345 (2005);
\bibitem{ryzhi}
 V. Ryzhii and V. Vyurkov, Phys. Rev. B {\bf 68} 165406
(2003); V. Ryzhii, Phys. Rev. B {\bf 68} 193402 (2003); V.Ryzhii
and R. Suris, J. Phys: Cond. Mat. 15, 6855, (2003) ; Ryzhii et al,
Sov. Phys. Semicond. 20, 1299, (1986).
\bibitem{rivera}
 P.H. Rivera
and P.A. Schulz, Phys. Rev. B {\bf 70} 075314 (2004)
\bibitem{shi}
 Junren Shi and
X.C. Xie, Phys. Rev. Lett. {\bf 91}, 086801 (2003).
\bibitem{ahn}
Kang-Hun Ahn, J. Korean Phys. Soc., {\bf 47} (4), 666-672, (2005).
\bibitem{andreev}
A.V. Andreev, I.L. Aleiner and A.J. Millis, Phys. Rev. Lett. {\bf
91}, 056803 (2003)
\bibitem{ina}
J. I\~narrea and G. Platero, Phys. Rev. Lett. {\bf 94} 016806,
(2005)
\bibitem{ng}
T-K Ng and Lixin Dai, Phys. Rev. B, {\bf 72}, 235333 (2005).
\bibitem{ina2}
J. I\~narrea and G. Platero, Phys. Rev. B {\bf 72} 193414 (2005)
\bibitem{ina3}
J.I\~narrea, G. Platero and C. Tejedor, Phys Rev. B. {\bf 50}
4581, (1994).
\bibitem{ina4}
J.I\~narrea and G. Platero  Phys Rev. B. {\bf 51} 5244, (1995).
\bibitem{kerner}
E.H. Kerner, Can. J. Phys. {\bf 36}, (3) 371-377 (1958) .
\bibitem{park}
K. Park, Phys. Rev. B {\bf 69} 201301(R) (2004).
\bibitem{ridley}
B.K. Ridley. Quantum Processes in Semiconductors, 4th ed. Oxford
University Press, (1993).


\end{thebibliography}
\end{document}